\documentclass{article} 
\usepackage{graphicx,float}
\usepackage{t1enc}
\usepackage[latin1]{inputenc}
\usepackage{mathrsfs,xspace,theorem}
\usepackage{amsmath,amssymb}
\usepackage{textcomp}
\usepackage{authblk} 
\theorembodyfont{\rmfamily}

\floatstyle{ruled}
\newfloat{Algorithm}{htpb}{alg}

\newcounter{prgline}

\newcounter{noqed}
\newcommand{\qed}{ \ifmmode\mbox{
}\fi\rule[-.05em]{.3em}{.7em}\setcounter{noqed}{0}}

\def\..{\,\mathpunct{\ldotp\ldotp}} % Middle stuff for intervals. Usage: \..

\newcommand{\url}{\cite{myurl}}

\newtheorem{lemma}{Lemma} 

\newtheorem{theorem}{Theorem}

\newtheorem{definition}{Definition}

\renewcommand{\epsilon}{\varepsilon}
\renewcommand{\phi}{\varphi}

\begin{document} 
\sloppy
\title{Succinct Dictionary Matching With No Slowdown}
\author{Djamal Belazzougui}
%\institute{LIAFA, Univ. Paris Diderot - Paris 7, 75205 Paris Cedex 13, France dbelaz@liafa.jussieu.fr}
\affil{LIAFA, Univ. Paris Diderot - Paris 7, 75205 Paris Cedex 13, France dbelaz@liafa.jussieu.fr}
%\date{}
\bibliographystyle{abbrv}
\maketitle
\begin{abstract}
The problem of dictionary matching is a classical problem in string matching: given a set $S$ of $d$ strings of total length $n$ characters over an (not necessarily constant) alphabet of size $\sigma$, build a data structure so that we can match in a any text $T$
all occurrences of strings belonging to $S$. The classical solution for this problem is the Aho-Corasick automaton which finds 
all $occ$ occurrences in a text $T$ in time $O(|T|+occ)$ using a data structure that occupies $O(m\log m)$ bits of space where $m\leq n+1$ is 
the number of states in the automaton. In this paper we show that the Aho-Corasick automaton can be represented in just $m(\log \sigma+O(1))+O(d\log(n/d))$ bits of space while still maintaining the ability to answer to queries in $O(|T|+occ)$ time. 
To the best of our knowledge, the currently fastest succinct data structure for the dictionary matching problem uses space $O(n\log \sigma)$ while answering queries in $O(|T|\log\log n+occ)$ time. In this paper we also show how the space occupancy can be reduced to $m(H_0+O(1))+O(d\log(n/d))$ where $H_0$ is the empirical entropy of the characters appearing in the trie representation of the set $S$, provided that $\sigma<m^\epsilon$ for any constant $0<\epsilon<1$. The query time remains unchanged. 
\end{abstract}
\section{introduction}
A recent trend in text pattern matching algorithms has been to succinctly encode data structures so that they occupy
no more space than the data they are built on, without a too significant sacrifice in their query time. The most prominent example being the data structures used for indexing texts for substring matching queries \cite{S00,FM00,GV00}. 
\\
In this paper we are interested in the succinct encoding of data structures for the dictionary matching problem, which consists in the construction of a data structure on a set $S$ of $d$ strings (a dictionary) of total length $n$ over an alphabet of size $\sigma$ (wlog we assume that $\sigma\leq n$) so that we can answer to queries of the kind: find in a text $T$ all occurrences of strings belonging to $S$ if any. 
The dictionary matching problem has numerous applications including computer security (virus detection software, intrusion detection systems), genetics and others. The classical solution to this problem is the Aho-Corasick automaton \cite{AC75}, which uses space $O(m\log m)$ bits (where $m$ is the number of states in the automaton which in the worst case equals $n+1$) and answers queries in time $O(|T|+occ)$ (where $occ$ is number of occurrences) if hashing techniques are used, or $O(|T|\log\sigma+occ)$ if only binary search is permitted. The main result of our paper is that the Aho-corasick automaton can be represented in just $m(\log\sigma+3.443+o(1))+d(3\log(n/d)+O(1))$ bits of space while still maintaining the same $O(|T|+occ)$ query time. As a corollary of the main result, we also show a compressed representation suitable for alphabets of size $\sigma<m^\epsilon$ for any constant $0<\epsilon<1$. This compressed representation uses $m(H_0+3.443+o(1))+O(d\log(n/d))$ bits of space where $H_0$ is the empirical entropy of the characters appearing in the trie representation of the set $S$. The query time of the compressed representation is also $O(|T|+occ)$.
\\
The problem of succinct encoding for dictionary matching has already been explored in \cite{CHLS05,HLSTV08,TWLY09,HLSTV09}. The results in \cite{CHLS05} and \cite{HLSTV09} only deal with the dynamic case which is not treated in this paper. The best results for the static case we have found in the literature are the two results from \cite{HLSTV08} and the result from \cite{TWLY09}. A comparison of the results from \cite{HLSTV08,TWLY09} with our main result is summarized in table \ref{table:compar_table}. In this table, the two results from \cite{HLSTV08} are denoted by HLSTV1  and HLSTV2 and the result of \cite{TWLY09} is denoted by TWLY. For comparison purpose, we have replaced $m$ with $n$ in the space and time bounds of our data structure. 

%\begin{figure}[htb]
%\centering
%\includegraphics[width=8cm]{auto.pdf}
%\caption{A split with a breadth first search starting in $r.$ Vertices 
%at distance $h$ compose $G[h].$ A connected compone
%nt of $G[>h]$ is marked $C.$}
%\label{gensplit}
%\end{figure}

\begin{table}
\centering
\label{table:compar_table}
\begin{tabular}{|l|l|l|}
  \hline
  Algorithm & Space usage (in bits) & Query time \\
  \hline
  HLSTV1 & $O(n\log\sigma)$ &$O(|T|\log\log(n)+occ)$ \\
  HLSTV2 & $n\log\sigma(1+o(1))+O(d\log(n))$ &$O(|T|(\log^{\epsilon}(n)+\log(d))+occ)$  \\
  TWLY & $n\log\sigma(2+o(1))+O(d\log(n))$ &$O(|T|(\log(d)+\log\sigma)+occ)$ \\
  Ours &$n(\log\sigma+3.443+o(1))+O(d\log(n/d))$ & $O(|T|+occ)$\\ 
\hline
\end{tabular}
\caption{Comparison of dictionary matching succinct indexes}
\end{table}

Our results assume a word RAM model, in which usual operations including multiplications, divisions and shifts are all supported in constant time. We assume that the computer word is of size $\Omega(\log n)$, where $n$ is the total size of the string dictionary on which we build our data structures. Without loss of generality we assume that $n$ is a power of two. All logarithms are intended as base $2$ logarithms.
We assume that the strings are drawn from an alphabet of size $\sigma$, where $\sigma$ is not necessarily constant. That is, $\sigma$ could be as large as $n$.
\\
The paper is organized as follows: in section \ref{sec:basic_components}, we present the main tools that will be used in our construction. 
%In section \ref{sec:aho_corasick} we do a brief recall of the Aho-Corasick automaton. 
In section \ref{sec:data_structure} we present our main result. In section \ref{sec:compressed_aho_corasick}, we give a compressed variant of the data structure. Finally some concluding remarks are given in section \ref{sec:conclusion}.
\section{Basic components}
\label{sec:basic_components}
In this paper, we only need to use three basic data structures from the literature of succinct data structures. 
\subsection{Compressed integer arrays}
We will use the following result about compression of integer arrays: 
\begin{lemma}
Given an array $A$ of $n$ integers such that $\sum_{0\leq i<n}A[i]=U$. We can produce a compressed representation that uses $n(\lceil\log(U/n)\rceil+2)$ bits of space such that any element of the array $A$ can be reproduced in constant time.   
\end{lemma}
This result was first described in \cite{GV00} based on Elias-Fano coding by Elias \cite{El74} and Fano \cite{F71} combined with succinct bitvectors \cite{CM96} which support constant time queries.  

\subsection{Succinctly encoded ordinal trees}
In the result of \cite{MR01} a tree of $n$ nodes of arbitrary degrees where the nodes are ordered in depth first order can be represented in $n(2+o(1))$ bits of space so that basic navigation on the tree can be done in constant time. %The result in \cite{MR01} uses a sequence of balanced parentheses where each node is represented by a pair of opening and closing parentheses. Navigation operations are implemented using basic operations on the parentheses sequence. Subsequently many other data structures were proposed to extend the set of supported navigation operations (the reader can refer to \cite{M09} for a survey of the main results in this area) in constant time. 
In this work we will only need a single primitive: given a node $x$ of preorder $i$ (the preorder of a node is the number attributed to the node in a DFS lexicographic traversal of the tree), return the preorder $j$ of the parent of $x$. 
%This primitive is supported by most of the succinct tree representations. 
\\The following lemma summarizes the result which will be used later in our construction:
\begin{lemma}
\label{lemma:succ_tree_lemma}
A tree with $n$ nodes of arbitrary degrees can be represented in $n(2+o(1))$, so that the preorder of the parent of a node of given preorder can be computed in constant time.   
\end{lemma}

In this paper we also use the compressed tree representation presented in \cite{JSS07} 
%(which is a compressed version of the so-called DFUDS representation presented in \cite{BDMRRR05}) 
which permits to use much less space than $2n+o(n)$ bits in the case where tree nodes degrees distribution is skewed (e.g. the tree has much more leaves than internal nodes). %This result supports many operations on compressed trees in constant time including the parent primitive.  
\begin{lemma}
\label{lemma:compr_tree_lemma}
A tree with $n$ nodes of arbitrary degrees can be represented in $n(H^*+o(1))$, where $H^*$ is the entropy of the degree distribution of the tree, so that the preorder of the parent of a node of given preorder can be computed in constant time.   
\end{lemma}

\subsection{Succinct indexable dictionary} 
In the paper by Raman, Raman and Rao \cite{RRR02} the following result is proved :
\begin{lemma}
\label{lemma:succ_idx_dict_lemma}
a dictionary on a set $\Gamma$ of $m$ integer keys from a universe of size $U$ can be built in time $O(m)$ and uses $B(m,U)+o(m)$ bits of space, where $B=\log\binom Um$ , so that the following two operations can be supported in constant time:
\begin{itemize}
\item $select(i)$: return the key of rank $i$ in lexicographic order (natural order of integers). 
\item $rank(k)$: return the rank of key $k$ in lexicographic order if $k\in \Gamma$. Otherwise return $-1$. 
\end{itemize}
\end{lemma}
The term $B=\log\binom Um$ is the information theoretic lower bound on the number of bits needed to encode all possible subsets of size $n$ of a universe of size $U$(we have $\binom Um$ different subsets and so we need $\log\binom Um$ to encode them). The term $B$ can be upper bounded in the worst case by $m(\log(e)+\log(U/m))$. The space usage of the dictionary can be simplified as $B(m,U)+o(m)\leq m(\log(e)+\log(U/m)+o(1))\leq m(\log(U/m)+1.443+o(1))$.
\section{The data structure} 
\label{sec:data_structure}
Given a set of strings $S$, our Aho-Corasick automaton has $|P|$ states where $P$ is the set of prefixes of strings in $S$. Each state of the automaton uniquely corresponds to one of the elements of $P$. Our Aho-Corasick representation has three kinds of transitions: $next$,$failure$ and $report$ transitions (the reader can refer to appendix~\ref{sec:aho_corasick} for a more detailed description of the transitions of the Aho-Corasick automaton). 
Our new data structure is very simple. We essentially use two succinctly encoded dictionaries, two succinctly encoded ordinal trees and one Elias-Fano encoded array. The representation we use is implicit in the sense that the strings of the dictionary are not stored at all. A query will output the occurences as triplets of the form $(occ\mathunderscore start\mathunderscore pos,occ\mathunderscore end\mathunderscore pos,string\mathunderscore id)$ where  $string\mathunderscore id$ is the identifier of a matched string from $S$ and $occ\mathunderscore start\mathunderscore pos$ ($occ\mathunderscore end\mathunderscore pos$) is the starting (ending) position of the occurrence in the text.   
\\
The central idea is to represent each state corresponding to a prefix $p\in P$, by a unique number $rank_P(p)\in[0,m-1]$ which represents the rank of $p$ in $P$ in suffix-lexicographic order (the suffix-lexicographic order is similar to lexicographic order except that the strings are compared in right-to-left order instead of left-to-right order). Then it is easy to see that the \emph{failure} transitions form a tree rooted at state $0$ (which we call a failure tree) and a DFS traversal of the tree will enumerate the states in increasing order. Similarly, the set of \emph{report} transitions represent a forest of trees, which can be transformed into a tree rooted at state $0$ (which call a report tree) by attaching all the roots of the forest as children of state $0$. Then similarly a DFS traversal of the report tree will also enumerate the states of the automaton in order. Then computing a \emph{failure} (\emph{report}) transition for a given state amounts to finding the parent of the state in the failure (report) tree. It turns out that the succinct tree representations (lemma \ref{lemma:succ_tree_lemma} and lemma \ref{lemma:compr_tree_lemma}) do support parent queries on DFS numbered trees in constant time. 

\subsection{State representation}
We now describe the state representation and the correspondence between states and strings. The states of our Aho-Corasick automaton representation are defined in the following way: 
\begin{definition}
Let $P$ be the set of all prefixes of the strings in $S$, and let $m=|P|$. We define the function $state$ as a function from $P$ into the interval $[0,m-1]$ where $state(p)$ is the rank of the string $p$ in $P$ according to the suffix-lexicographic order(we count the number of elements of $P$ which are smaller than $p$ in the suffix lexicographic order). 
\label{states_definition}
\end{definition}
The suffix-lexicographic order is defined in the same way as standard lexicographic order except that the characters of the strings are compared in right-to-left order instead of left-to-right order. That is the strings of $P$ are first sorted according to their last character and then ties are broken according to their next-to-last character, etc\ldots. In order to distinguish final states from the other states, we simply note that we have exactly $d$ terminal states corresponding to the $d$ elements of $S$. As stated in the definition, each of the $m$ states is uniquely identified by a number in range $[0,m-1]$. Therefore in order to distinguish terminal from non-terminal states, we use a succinct indexable dictionary, in which we store the $d$ numbers corresponding to the $d$ terminal states. As those $d$ numbers all belong to the range $[0,m-1]$, the total space occupation of our dictionary is $d(\log(m/d)+1.443+o(1))$ bits. In the following, we denote this dictionary as the state dictionary. 

\subsection{Representation of next transitions}
We now describe how $next$ transitions are represented. First, we note that a transition goes always from a state corresponding to a prefix $p$ where $p\in P$ to a state corresponding to a prefix $pc$ for some character $c$ such that $pc\in P$. Therefore in order to encode the transition labeled with character $c$ and which goes from the state corresponding to the string $p$ (if such transition exists), we need to encode two informations: whether there exists a state corresponding to the prefix $pc$ and the number corresponding to that state if it exists. In other words, given $state(p)$ and a character $c$, we need to know whether there exists a state corresponding to $pc$ in which case, we would wish to get the number $state(pc)$. 
\\
The transition from $state(p)$ to $state(pc)$ can be done in a very simple way using a succinct indexable dictionary (lemma \ref{lemma:succ_idx_dict_lemma}) which we call the transition dictionary. For that we notice that $state(p)\in [0,m-1]$. For each non empty string $p_i=p'_ic_i$ where $p_i\in P$, we store in the transition dictionary, the pair $pair(p_i)=(c_i,state(p'_i))$ as the concatenation of the bit representation of $c_i$ followed by the bit representation of $state(p'_i)$. That is we store a total of $m-1$ pairs which correspond to the $m-1$ non empty strings in $P$. Notice that the pairs are from a universe of size $\sigma m$. Notice also that the pairs are first ordered according the characters $c_i$ and then by $state(p'_i)$ (in the $C$ language notation a pair is an integer computed as $pair(p_i)=(c_i<<\log m)+state(p'_i)$. Now the following facts are easy to observe: 
\begin{enumerate}
\item Space occupation of the transition dictionary is $m(\log((\sigma \cdot m)/m)+1.443+o(1))=m(\log \sigma+1.443+o(1))$. 
\item The rank of the pairs stored in the succinct dictionary reflects the rank of the elements of $P$ in suffix-lexicographic order. This is easy to see as we are sorting pairs corresponding to non empty strings, first by their last characters before sorting them by the rank of their prefix excluding their last character. Therefore we have $rank(pair(p_i))=state(p_i)+1$, where $rank$ function is applied on the transition dictionary. 
\item A pair $(c_i,state(p'_i))$ exists in the transition dictionary if and only if we have a transition from the state corresponding to $p'_i$ to the state corresponding to $p'_ic_i$ labeled with the $c_i$. 
\end{enumerate} 
From the last two observations we can see that a transition from a state $state(p)$ for a character $c$ can be executed in the following way: first compute the pair $(c,state(p))$. Then query the transition dictionary using the function $rank((c,state(p)))$. If that function returns $-1$, we can deduce that there is no transition from $state(p)$ labeled with character $c$. Otherwise we will have $state(pc)=rank((c,state(p)))+1$. In conclusion we have the following lemma:
\begin{lemma}
\label{lemma:next_trans_lemma}
The $next$ transitions of an Aho-corasick automaton whose states are defined according to definition~\ref{states_definition} can be represented in $m(\log\sigma+1.443+o(1))$ bits of space such that the existence and destination state of a transition can be computed in constant time.  
\end{lemma}

\subsection{Representation of failure transitions}
We now describe how failure transitions are encoded. Recall that a failure transition connects a state representing a prefix $p$ to the state representing $q$ where $q$ is the longest suffix of $p$ such that $q\in P$ and $q\neq p$. The set of failure transitions can be represented with a tree called the \emph{failure tree}. Each node in the failure tree represents an element of $P$ and each element of $P$ has a corresponding node in the tree. The failure tree is simply defined in the following way: 
\begin{itemize}
\item The node representing a string $p$ is a descendant of a node representing the string $q$ if and only if $q\neq p$ and $q$ is suffix of $p$.
\item The children of any node are ordered according to the suffix-lexicographic order of the strings they represent.
\end{itemize}

%\\ %We now briefly describe how to build the tree conceptually. The root of the tree represents the empty string. We think of the levels of the tree as numbered from $0$ from top to bottom. At level $0$, we put only the empty string. We note the set containing only the empty string by $P_0$. At level $1$, we put all non empty strings of $R_1=P-P_0$ that have no suffix in $R_1=P-P_0$ except for the empty string. Those nodes are all attached as children of the root and ordered in increasing suffix-lexicographic order of the strings they represent. Now at level $2$ of the tree we put the set $P_2$ of strings of $R_2=R_1-P_1=P-(P_0\cup P_1)$ that have no suffix in $R_2$. We attach the node representing each string of $P_2$ as a child of the only node at level $1$ representing a string which is a suffix of $P_2$. For any node at level $1$ (which represent an element of $P_1$), we order all its children (which represent elements from $P_2$ ) in suffix-lexicographic order. More generally, at any level $i$, we will put elements of $R_i=R_{i-1}-P_{i-1}$ that have no suffix in $R_i$. We attach as children for any element $p$ at level $i$ all elements at level $i+1$ that have $p$ as a suffix and we order them in suffix-lexicographic order(strings are compared character by character from right to left).
%\\
Now an important observation on the tree we have just described is that a depth first traversal of the tree will enumerate all the elements of $P$ in suffix-lexicographic order. That is the preorder of the nodes in the tree corresponds to the suffix lexicographic order of the strings of $P$. It is clear from the above description that finding the failure transition that connects a state $state(p)$ to a state $state(q)$ (where $q$ is the longest element in $P$ such that $q$ is a suffix of $p$ and $q\neq p$) corresponds to finding the parent in the failure tree of the node representing the element $q$. 
Using a succinct encoding (lemma \ref{lemma:succ_tree_lemma}), the tree  can be represented using space $2m+o(m)$ bits such that the parent primitive is supported in constant time. That is the node of the tree corresponding to a state $p$ will have preorder $state(p)$, and the preorder of the parent of that node is $state(q)$. A failure transition is computed in constant time by $state(q)=parent(state(p))$. 
\begin{lemma}
\label{lemma:fail_trans_lemma}
The failure transitions of the Aho-corasick automaton whose states are defined according to definition~\ref{states_definition} can be represented in $m(2+o(1))$ bits of space such that a failure transition can be computed in constant time.
\end{lemma}
\subsection{Representation of report transitions}
The encoding of the $report$ transitions is similar to that of failure transitions. The only difference with the failure tree is that any non root internal node is required to represent an element of $S$. We remark that the report transitions form a forest of trees, which can be transformed into a tree by connecting all the roots of the forest as children of state $0$. In other words a report tree is the unique tree built on the elements of $P$ which satisfies :
\begin{itemize} 
\item All the nodes of the tree represent strings of $P$.
\item All the nodes are descendants of the root which represents the empty string.
\item The node representing a string $p$ is a descendant of a node representing a non empty string $s$ if and only if $s\in S$, $s\neq p$ and $s$ is a suffix of $p$. 
\item All children of a given node are ordered according to the suffix-lexicographic order of the strings they represent.
\end{itemize}

We could encode the report tree in the same way as the failure tree (using lemma \ref{lemma:succ_tree_lemma}) to occupy $m(2+o(1))$ bits of space. However we can obtain better space usage if we encode the report tree using the compressed tree representation (lemma \ref{lemma:compr_tree_lemma}). More specifically, the report tree contains at most $d$ internal nodes as only strings of $S$ can represent internal nodes. This means that the tree contains at least $m-d$ leaves. The entropy of the degree distribution of the report tree is $d(\log(m/d)+O(1))$ bits and the encoding of lemma \ref{lemma:compr_tree_lemma} will use that much space (this can easily be seen by analogy to suffix tree representation in \cite{JSS07} which uses $d(\log((d+t)/d)+O(1))$ bits of space for a suffix tree with $d$ internal nodes and $t$ leaves). Report transitions are supported similarly to failure transitions in constant time using the parent primitive which is also supported in constant time by the compressed tree representation.% of \cite{JSS07}.  
\begin{lemma}
\label{lemma:rep_trans_lemma}
The report transitions of the Aho-corasick automaton whose states are defined according to definition~\ref{states_definition} can be represented in $d(\log(m/d)+O(1))$ bits of space such that a report transition can be computed in constant time.
\end{lemma}
\subsection{Occurence representation}
Our Aho-corasick automaton will match strings from $S$ which are suffixes of prefixes of the text $T$. This means that the Aho-corasick automaton will output the end positions of occurrences. However the user might need to also have the start position of occurrences. For that we have chosen to report occurrences as triplets $(occ\mathunderscore start\mathunderscore pos,occ\mathunderscore end\mathunderscore pos,string\mathunderscore id)$, where $string\in S$ and  $occ\mathunderscore start\mathunderscore pos$ ($occ\mathunderscore end\mathunderscore pos$) is the start (end) position of the occurrence in the text. For that we need to know the length of the matched strings. But this information is not available as we do not store the original strings of the dictionary in any explicit form. For that purpose, we succinctly store an array storing the dictionary string lengths using the Elias-Fano encoding and call the resulting compressed array, a string length store. As the total length of the strings of the dictionary is $n$, the total space usage of Elias-Fano encoded array will be $d(\lceil\log(n/d)\rceil+2)$. 
\\
We note that our algorithm outputs string identifiers as numbers from interval $[0,d-1]$ where the identifier of each string corresponds to the rank of the string in the suffix lexicographic order of all strings. If the user has to associate specific action to be applied when a given string is matched, then he may use a table $action$ of size $d$ where each cell $i$ stores the value representing the action associated with the string of rank $i$ in suffix lexicographic order. The table could be sorted during the building of the state dictionary. 
\subsection{Putting things together}
Summarizing the space usage of the data structures which are used for representation of the Aho-Corasick automaton: 
\begin{enumerate}
\item The state dictionary which indicates the final states occupies at most $d(\log(m/d)+1.443+o(1))\leq d(\log(n/d)+1.443+o(1))$ bits of space. 
\item The $next$ transitions representation occupies $B(m,m\sigma)+o(m)\leq m(\log(\sigma)+1.443+o(1))$ bits of space. 
\item The $failure$ transitions representation occupies $m(2+o(1))$ bits of space.
\item The $report$ transitions representation occupies $d(\log(m/d)+O(1))\leq d(\log(n/d)+O(1))$ bits of space. 
\item The string length store occupies $d(\lceil\log(n/d)\rceil+O(1))$ bits of space.
\end{enumerate}
The following lemma summarizes the space usage of our representation: 
\begin{lemma}
\label{lemma:aho_space_lemma}
The Aho-corasick automaton can be represented in $m(\log\sigma+3.443+o(1))+d(3\log(n/d)+O(1))$ bits of space.
\end{lemma}
A more detailed space usage analysis is left to appendix~\ref{sec:space_analysis}.
\\
\paragraph{Implicit representation of the dictionary strings}
We note that the state dictionary and the transition dictionary can be used in combination as an implicit representation of the elements of $S$. 
\begin{lemma}
\label{lemma:implicit_dict_lemma}
For any integer $i\in[0,d-1]$, we can retrieve the string $x\in S$ of rank $i$ (in suffix-lexicographic order) in time $O(|x|)$ by using the transition dictionary and state dictionary. 
\end{lemma}
The proof of the lemma is left to appendix~\ref{implicit_dict}. 

%\begin{figure} 
%\caption{Report tree for the the set \{"ABC","B","BC","CA"\}} %la légende
%\label{report_tree_drawing} %l'étiquette pour faire référence à cette image
%\end{figure} %on ferme l'environnement figure

\subsection{Queries}
Our query procedure essentially simulates the Aho-Corasick automaton operations, taking a constant time for each simulated operation. Thus our query time is within a constant factor of the query time of the original Aho-Corasick. 
\begin{lemma}
\label{lemma:aho_time_lemma}
The query time of the succinct Aho-Corasick automaton on a text $T$ is $O(|T|+occ)$, where $occ$ is the number of reported occurrences.
\end{lemma}
Details of the query algorithm are left to appendix~\ref{sec:query_algorithm}.
\subsection{Construction}
We now describe the construction algorithm which takes $O(n)$ time . The algorithm is very similar to the one described in~\cite{DL05}. 
We first write each string $s_i$ of $S$ in reverse order and append a special character $\#$ at the end of each string giving a set $R$. 
The character $\#$ is considered as smaller than all characters of original alphabet $\sigma$. 
Then, we build a (generalized) suffix-tree on the set $R$. This can be done in time $O(n)$ using the algorithm in \cite{F97} for example. Each leaf in the suffix tree will store a list of suffixes where a suffix $s$ of a string $x\in R$ is represented by the pair $(string\mathunderscore pointer,suf\mathunderscore pos)$, where $string\mathunderscore pointer$ is a pointer to $x$ and $suf\mathunderscore pos$ is the starting position of $s$ in $x$.
Then we can build the following elements:
\begin{enumerate}
\item The transition dictionary can be directly built as the suffix tree will give us the (suffix-lexicographic) order of all elements of $P$ by a DFS traversal (top-down lexicographic traversal).
\item The failure tree is built by a simple DFS traversal of the suffix tree.
\item The report tree is built by doing a DFS traversal of the failure tree.
\item The state dictionary can be built by a traversal of the report tree.
\item The string length store can be built by a simple traversal of the set $S$. 
\end{enumerate}
Details of the construction are left to Appendix~\ref{sec:construction_details}. 
\begin{lemma}
\label{lemma:aho_build_lemma}
The succinct Aho-corasick automaton representation can be constructed in time $O(n)$. 
\end{lemma}
The results about succinct Aho-Corasick representation are summarized by the following theorem: 
\begin{theorem}
\label{main_theorem}
The Aho-corasick automaton for a dictionary of $d$ strings of total length $n$ characters over an alphabet of size $\sigma$ can be represented in $m(\log\sigma+3.443+o(1))+d(3\log(n/d)+O(1))$ bits where $m\leq n+1$ is the number of states in the automaton. A dictionary matching query on a text $T$ using the Aho-corasick representation can be answered in $O(|T|+occ)$ time, where $occ$ is the number of reported strings. The representation can be constructed in $O(n)$ randomized expected time. 
\end{theorem}
\section{Compressed representation}
\label{sec:compressed_aho_corasick}
The space occupany of theorem \ref{main_theorem} can be further reduced to $m(H_0+3.443+o(1))+d(3\log(n/d)+O(1))$, where $H_0$ is the entropy of the characters appearing as labels in the $next$ transitions of the Aho-Corasick automaton.
\begin{theorem}
\label{compressed_theorem}
The Aho-corasick automaton for a set $S$ of $d$ strings of total length $n$ characters over an alphabet of size $\sigma$ can be represented in $m(H_0+3.443+o(1))+d(3\log(n/d)+O(1))$ bits where $m\leq n+1$ is the number of states in the automaton and $H_0$ is the entropy of the characters appearing in the trie representation of the set $S$. The theorem holds provided that $\sigma<m^\epsilon$ for any constant $0<\epsilon<1$. A dictionary matching query for a text $T$ can be answered in $O(|T|+occ)$ time. 
\end{theorem}
The proof of the theorem is deferred to appendix~\ref{compressed_aho}.
\section{Concluding remarks}
\label{sec:conclusion}

%We have shown a simple construction of a succinct Aho-Corasick representation achieving space $m(\log\sigma+3.443+o(1))+O(d\log(n/d))$ bits where $m$ is the number of states in the Aho-Corasick automaton. This is much better than the $O(m \log m)$ bits used by the original Aho-Corasick representation. Query time of our representation is $O(|T|+occ)$ which is the same as that of the non-succinct Aho-Corasick. Our data structure can be constructed in  $O(n)$ time using a construction algorithm similar to the one described in paper \cite{DL05}. Compared with the other succinct dictionary matching data structures we found in the literature, our construction achieves better query time than the fastest succinct construction which incurs s $\log\log n$ factor slowdown. The space usage achieved by our data structure although optimal up to an additive $O(n)$ term is worse than the other succinct representations in case the alphabet is of (small) constant size. However for super-constant alphabet sizes ($\sigma=\Omega(\log n)$ for example) the space usage of our data structure is as good as the most compact succinct data structure  (second result in \cite{HLSTV08}). \\
Our work gives rise to two open problems: the first one is whether the term $3.443m$ in the space usage of our method which is particularly significant for small alphabets (DNA alphabet for example) can be removed without incurring any slowdown. The second one is whether the query time can be improved to $O(|T|\log\sigma/w+occ)$ (which is the best query time one could hope for). 
%The naive solution which consists in merging multiple transitions of the Aho-Corasick automaton would not work as it would result in an explosion of the number of transitions.
%\item can the $m\log\sigma$ term in space usage be reduced to $nH_k$ (where $nH_k$ denotes the $k$th order entropy) while still maintaining the same time bound. The method described in paper \cite{HLSTV08} uses $n(H_k+o(\log\sigma))+O(d\log n)$, but has a slowdown factor $\log^{\epsilon}n+\log d$. 
%We note that the Aho-Corasick representation 
  
%\end{itemize}
\section*{Acknowledgements}
The author is grateful to Mathieu Raffinot for proofreading the paper and for useful comments and suggestions. The author wishes to thank Kunihiko Sadakane and Rajeev Raman for confirming that the construction time of their respective data structures in \cite{JSS07} and \cite{RRR02} is linear.
\small 
\bibliography{Aho-corasick} 

\begin{thebibliography}{10}

\bibitem{AC75}
A.~V. Aho and M.~J. Corasick.
\newblock Efficient string matching: An aid to bibliographic search.
\newblock {\em Commun. ACM}, 18(6):333--340, 1975.

\bibitem{CHLS05}
H.-L. Chan, W.-K. Hon, T.~W. Lam, and K.~Sadakane.
\newblock Dynamic dictionary matching and compressed suffix trees.
\newblock In {\em SODA}, pages 13--22, 2005.

\bibitem{CM96}
D.~R. Clark and J.~I. Munro.
\newblock Efficient suffix trees on secondary storage (extended abstract).
\newblock In {\em SODA}, pages 383--391, 1996.

\bibitem{DL05}
S.~Dori and G.~M. Landau.
\newblock Construction of aho corasick automaton in linear time for integer
  alphabets.
\newblock In {\em CPM}, pages 168--177, 2005.

\bibitem{El74}
P.~Elias.
\newblock Efficient storage and retrieval by content and address of static
  files.
\newblock {\em J. ACM}, 21(2):246--260, 1974.

\bibitem{F71}
R.~M. Fano.
\newblock On the number of bits required to implement an associative memory.
\newblock Memorandum 61, Computer Structures Group, Project MAC, MIT,
  Cambridge, Mass., n.d., 1971.

\bibitem{F97}
M.~Farach.
\newblock Optimal suffix tree construction with large alphabets.
\newblock In {\em FOCS}, pages 137--143, 1997.

\bibitem{FM00}
P.~Ferragina and G.~Manzini.
\newblock Opportunistic data structures with applications.
\newblock In {\em FOCS}, pages 390--398, 2000.

\bibitem{GV00}
R.~Grossi and J.~S. Vitter.
\newblock Compressed suffix arrays and suffix trees with applications to text
  indexing and string matching (extended abstract).
\newblock In {\em STOC}, pages 397--406, 2000.

\bibitem{HLSTV08}
W.-K. Hon, T.~W. Lam, R.~Shah, S.-L. Tam, and J.~S. Vitter.
\newblock Compressed index for dictionary matching.
\newblock In {\em DCC}, pages 23--32, 2008.

\bibitem{HLSTV09}
W.-K. Hon, T.~W. Lam, R.~Shah, S.-L. Tam, and J.~S. Vitter.
\newblock Succinct index for dynamic dictionary matching.
\newblock In {\em ISAAC}, 2009.

\bibitem{JSS07}
J.~Jansson, K.~Sadakane, and W.-K. Sung.
\newblock Ultra-succinct representation of ordered trees.
\newblock In {\em SODA}, pages 575--584, 2007.

\bibitem{MR01}
J.~I. Munro and V.~Raman.
\newblock Succinct representation of balanced parentheses and static trees.
\newblock {\em SIAM J. Comput.}, 31(3):762--776, 2001.

\bibitem{RRR02}
R.~Raman, V.~Raman, and S.~S. Rao.
\newblock Succinct indexable dictionaries with applications to encoding k-ary
  trees and multisets.
\newblock In {\em SODA}, pages 233--242, 2002.

\bibitem{S00}
K.~Sadakane.
\newblock Compressed text databases with efficient query algorithms based on
  the compressed suffix array.
\newblock In {\em ISAAC}, pages 410--421, 2000.

\bibitem{TWLY09}
A.~Tam, E.~Wu, T.~W. Lam, and S.-M. Yiu.
\newblock Succinct text indexing with wildcards.
\newblock In {\em SPIRE}, pages 39--50, 2009.

\end{thebibliography}
\normalsize

\newpage

\appendix
\section{Original Aho-Corasick automaton}
\label{sec:aho_corasick}
\begin{figure}[htb] 
\centering\includegraphics[scale=0.6]{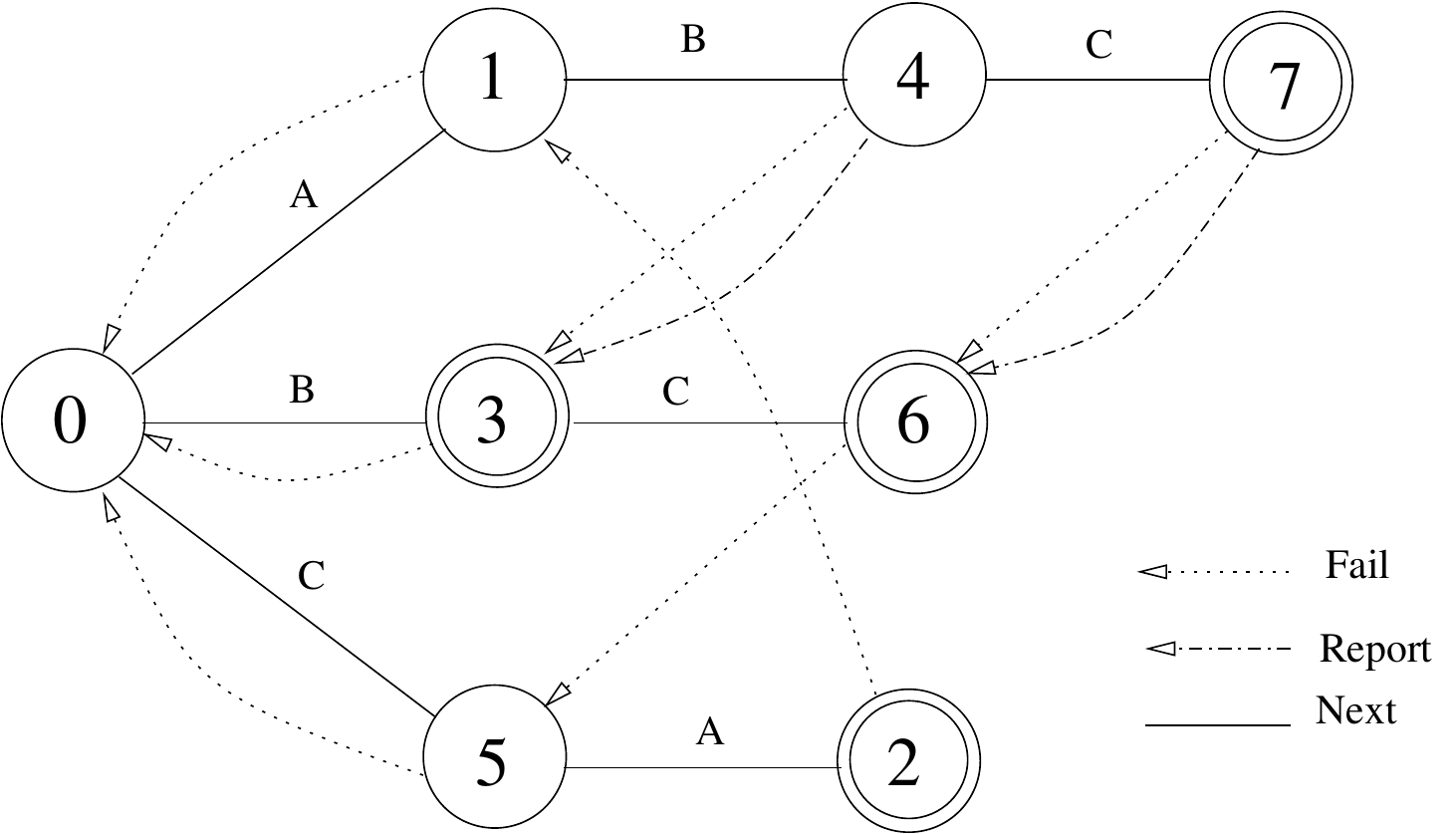} 
\caption{The Aho-Corasick automaton for the set \{"ABC","B","BC","CA"\}} %la légende
\label{automaton_drawing} %l'étiquette pour faire référence à cette image
\end{figure}

We recall the original Aho-Corasick automaton. The variant described here may slightly differ from other ones for the reason that this variant is simpler to adapt to our case. In particular the strings of $S$ are implicitly represented by the automaton and are never represented explicitly. 
\\
In the Aho-Corasick automaton, we have $m$ states with three kinds of transitions: $next$, $fail$ and $report$ transitions. We define the set $P$ as the set of all prefixes of the strings of the set $S$, including the empty string and all the strings of $S$. For each element of $P$, we have one corresponding state in the automaton and vice-versa. We thus have $|P|=m\leq n+1$ states in the automaton. The states that correspond to strings in $S$ are called terminal states.  We now define the three kinds of transitions: 
\begin{itemize}
\item For each state $v_{p}$ corresponding to a prefix $p$, we have a transition $next(v_{p},c)$ labeled with character $c$ from the state $v_p$ to each state $v_{pv}$  corresponding to a prefix $pc$ for each prefix $pc\in{P}$. Hence we may have up to $\sigma$ $next$ transitions per state. 
\item For each state $v_{p}$ we have a failure transition which connects $v_{p}$ to the state $v_q$ corresponding to the longest suffix $q\in{P}$ of $p$ such that $p\neq q$.
\item Additionally , for each state $v_{p}$, we may have a $report$ transition from the state $v_{p}$ to the state corresponding to the longest suffix $q$ of $p$ such that $q\in{S}$ and $p\neq q$ if such $q$ exists. If for a given state $v_{p}$ no such string exists, then we do not have a report transition from the state $v_p$. 
\end{itemize}
Figure~\ref{automaton_drawing} shows an example of an Aho-Corasick automaton with the three kinds of transitions.
\\
We now turn out to the algorithm which recognizes the patterns of $S$ included in a text $T$ using the Aho-Corasick automaton. The algorithm works in the following way: initially, before scanning the text the automaton is at the state $zero$ which corresponds to the empty string. Then at each step: 
\begin{itemize}
\item We first check if the string corresponding to the current state (which we note by $state_{cur}$) has any element of $S$ as a suffix. If this is the case we enumerate those elements in their decreasing length. This works in the following way: if $state_{cur}$ is a terminal state, then we report the string corresponding to $state_{cur}$ as a matching one. If the state has a $report$ transition, then we skip to the state corresponding to that transition (note that this state is necessarily a terminal state) and report the string corresponding to that state as a matching string. If that state has also a $report$ transition, we continue recursively following $report$ transitions and report the strings corresponding to the traversed states until we reach a state that has no report transition. 
\item We read the next character $c$ from the text. Then we determine the next state to skip to. To that effect, we first check if there exists a $next$ transition from $state_{cur}$ labeled with the character $c$. If this is the case, follow that transition and go to the state indicated by the transition. Otherwise, we have to follow the $fail$ transition and do the same thing recursively (check whether there exists a $next$ transition from that new state labeled with character $c$, etc$\ldots$). If at some point we reach the state $zero$ corresponding to the empty string and fail to find a next transition from it labeled with $c$, we stay in state $zero$.  
\end{itemize} 
 %on ferme l'environnement figure
\section{Proof of lemma~\ref{lemma:implicit_dict_lemma}}
\label{implicit_dict}
Retrieving the pattern $x$ of $S$ of rank $i$ (in suffix lexicographic order) can be done in the following way: we first do a $select(i)$ on the state dictionary giving a number $j_0$. Then by doing a $select(j_0)$ on the transition dictionary we obtain a pair $(c_1,j_1)$. The character $c_1$ is in fact the last character of the string we are looking for. Then we do a $select(j_1)$ obtaining a pair $(c_2,j_2)$ where $c_2$ is the second-to-last character and we continue that way until we obtain a pair $(c_{|x|},0)$, where $c_{|x|}$ is the first character of the string $x$. 
\\The justification for this is that a $select(i)$ will give us $j_0$, the rank of the string in $x$ in suffix lexicographic order, relatively to the set $P$. In other words $j_0$ is the terminal state corresponding to the string $x$. Then by doing a $select(j_0)$ we will obtain a pair $(c_1,j_1)$, where $j_1$ is the state corresponding to $y$ the prefix of $x$ of length $|x|-1$. The necessarily $c_1$ is the last character of $x$, because the pair $(c_1,j_1)$ corresponds to the transition which lead to the state corresponding to $x$. Continuing that way recursively doing $select$ queries with the states returned by preceding $select$ queries, we will report all characters of $x$ in right-to-left order. 

\section{Query algorithm}\label{sec:query_algorithm}
We now present the full picture of the queries. The text $T$ is scanned from left to right. We use a variable named $state$ which is initialized to zero, a temporary variable called $tmp\mathunderscore state$, a variable $dict\mathunderscore string\mathunderscore idx$ which stores the index of the reported string, a variable called $dict\mathunderscore string\mathunderscore length$ and finally a variable $step$ which stores the current step number (initially set to zero). At each step we do the following actions: 
\begin{enumerate}
\item Read the character $c=T[step]$ and increment variable $step$: $step=step+1$. 
\item Check the existence of a transition from the state $state$ labeled with character $c$. For that we probe the transition dictionary for the pair $(c,state)$ by doing a $tmp\mathunderscore state=rank((c,state))$ where rank is applied on the transition dictionary. Then if we have $tmp\mathunderscore state\neq -1$, we set $state=tmp\mathunderscore state$ and go to step 3. Otherwise we conclude that there exists no transition from the state $state$ labeled with character $c$. If $state=0$, this step is finished without matching any string of $S$.  Otherwise if $step\neq 0$, we go to step $3$ in order to do a failure transition.
\item Find the destination state of the failure transition from the state $state$. This corresponds to doing a parent operation on the succinctly encoded failure tree, which takes a constant time. We set the variable $state$ to the destination state and return to step $2$. 
\item We first set the variable $tmp\mathunderscore state=state$. Then check if the current state matches any string in the dictionary. For that we check whether $state$ is stored in the state dictionary. If this is the case we go to step 6, otherwise we go to step 5. 
\item We do a parent operation on the report tree for $tmp\mathunderscore state$ and store the returned state in variable $tmp\mathunderscore state$. If $tmp\mathunderscore state=0$, we conclude that there exists no report transition from state $tmp\mathunderscore state=state$ and thus return to step $1$ in order to process the next character in the text. 
\item We have now to report the string corresponding to state $tmp\mathunderscore state$ as a matching string. For that, we first do a $dict\mathunderscore string\mathunderscore idx=rank(tmp\mathunderscore state)$ where $rank$ operation is applied on the state dictionary. This operation returns a unique number corresponding to one of the strings of $S$. Then we retrieve the length of the string number $dict\mathunderscore string\mathunderscore idx$ from the dictionary string lengths store and store it in the variable $dict\mathunderscore string\mathunderscore length$. We finally report the occurrence as the triplet $(occ\mathunderscore start\mathunderscore pos,occ\mathunderscore end\mathunderscore pos,string\mathunderscore id)=(step-dict\mathunderscore string\mathunderscore length,step-1,dict\mathunderscore string\mathunderscore idx)$. At the end of the step we return to step $5$.
\end{enumerate}
\section{Detailed space analysis}\label{sec:space_analysis}
If we compare $m(\log\sigma+3.443+o(1))+d(3\log(n/d)+O(1))$, the space usage of our data structure with the lower bound $n\log\sigma$, which represents the total size of the patterns, then we can see the following facts : 
\begin{itemize}
\item The $B(m,m\sigma)+o(m)$ term is slightly smaller for small values of $\sigma$. For example, for $\sigma=2$, it equals $m(2+o(1))=m(\log(\sigma)+1)+o(m)$. 
\item The $cn$ term can be simplified to $o(n\log\sigma)$ for non constant $\sigma$. 
\item We have $d(3\log(n/d)+O(1))=o(n\log\sigma)$ for any value of $\sigma$. This can easily be seen: we have $d$ strings of total length $n\log\sigma$. The average length of the strings must be at least $(\log n)$ bits (otherwise the strings will not be all distinct). The space usage is maximized when the ratio $d/n$ is maximal. This happens at the point $n\log\sigma=d\log n$, where we have $d(3\log(n/d)+O(1))=d(3\log\log n+O(1))=(n\log\sigma)((3\log\log n+O(1))/\log n)=o(n\log\sigma)$.
\end{itemize}
The space usage for different values of $\sigma$ is summarized in table \ref{table:space_usage_table}, where the space optimality ratio refers to the ratio between actual space usage and the optimal $n\log\sigma$ bits of space. 
\begin{table}
\centering
\begin{tabular}{|l|l|l|}
  \hline
  $\sigma$ (alphabet size) & Space usage & Space optimality ratio \\
  \hline
  2 & $n(4+o(1))$ & 4\\
  4 & $n(5.246+o(1))$ & 2.623\\
  16 & $n(7.397+o(1))$ & 1.849\\
  256 & $n(11.440+o(1))$ & 1.43\\
  Superconstant & $n(\log\sigma+3.443+o(1))$ & 1\\
\hline
\end{tabular}
\caption{Space optimality ratio for different alphabet sizes}
\label{table:space_usage_table}
\end{table} 

\section{A full example}\label{sec:full_example}
\begin{figure}[htb]
\centering\includegraphics[scale=0.5]{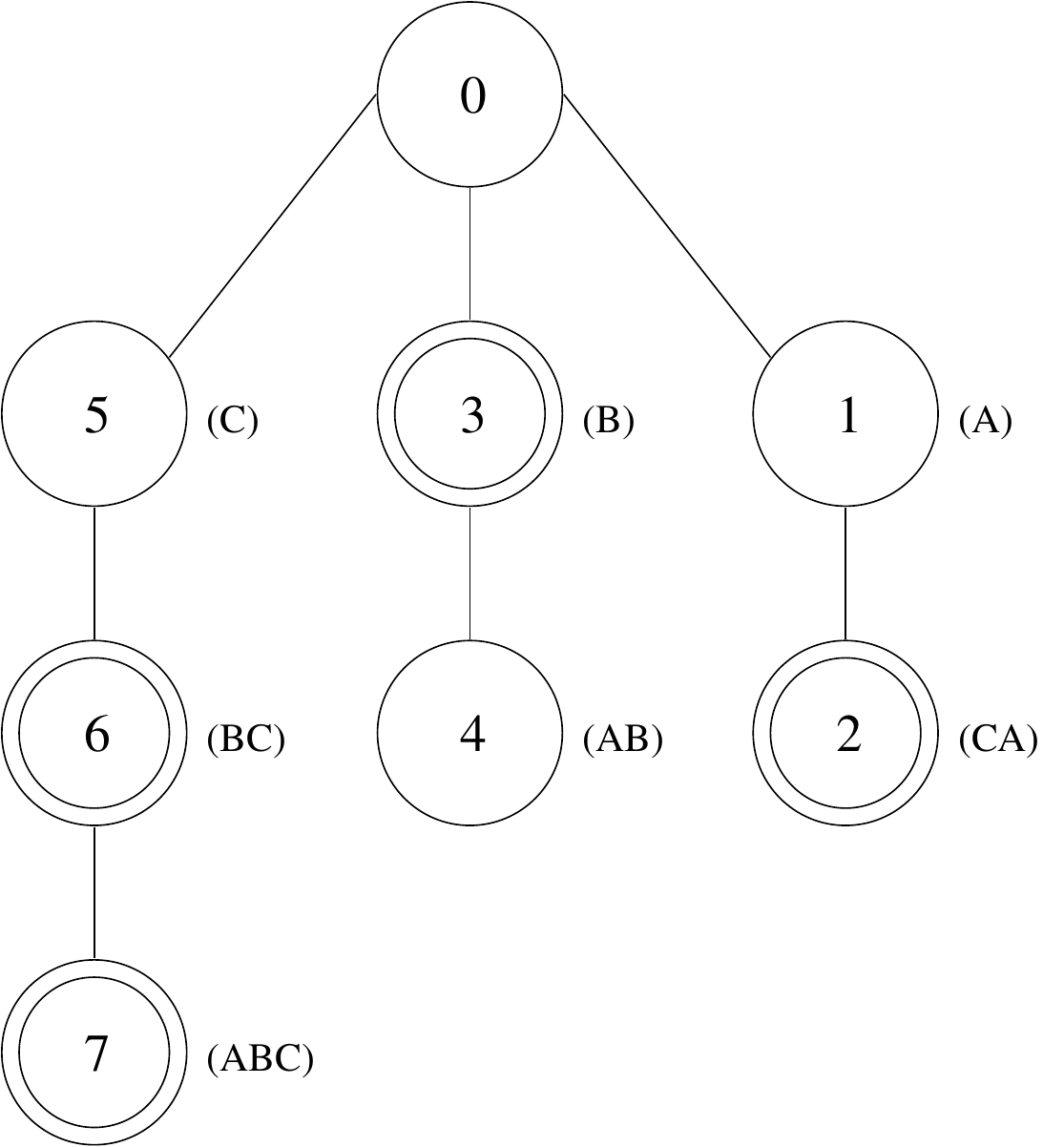} 
\includegraphics[scale=0.45]{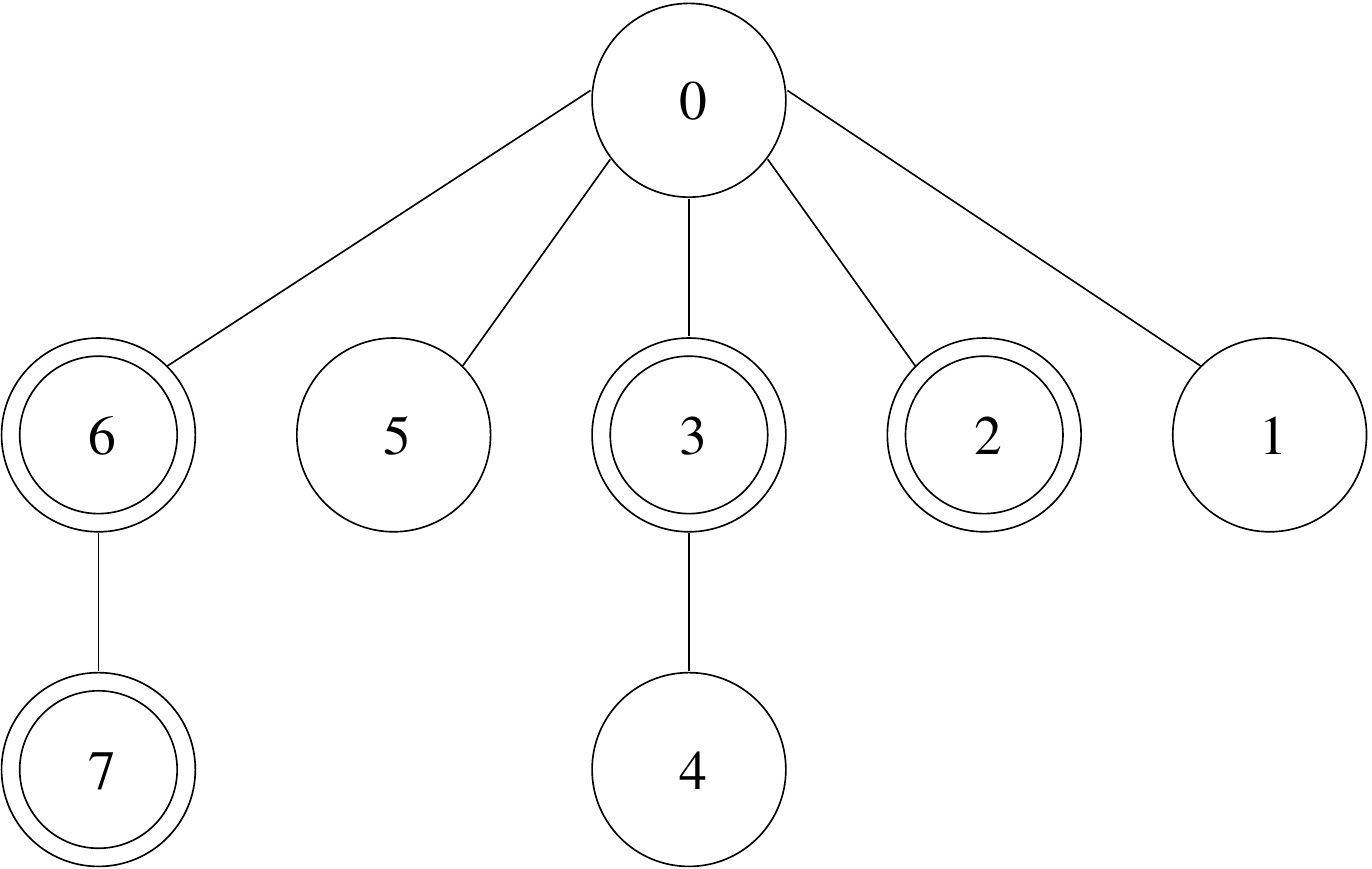}
\caption{Failure and report trees for the set \{"ABC","B","BC","CA"\}} %la légende
\label{failure_tree_drawing} %l'étiquette pour faire référence à cette image
\end{figure}

Let's now take as example the set $S=\{''ABC'',''B'',''BC'',''CA''\}$. The example is illustrated in figures \ref{automaton_drawing} and \ref{failure_tree_drawing}. The set $P$ of prefixes of $S$ sorted in suffix-lexicographic order gives the sequence: $''~'',~''A'',~''CA'',~''B'',~''AB'',~''C'',~''BC'',~''ABC''$. The first prefix (the empty string) in the sequence corresponds to state $0$ and the last one corresponds to state $7$. For this example, we store the following elements: 
\begin{itemize}
\item The transition dictionary stores the following pairs: $(A,0),(A,5),(B,0),(B,1),(C,0),(C,3),(C,4)$. 
\item The state dictionary stores the states $2,3,6,7$ which correspond to the final states of the automaton (states corresponding to the strings of $S$). 
\item The report and failure trees are depicted in figure \ref{failure_tree_drawing}.
\item The string length store, stores the sequence $2,1,2,3$ which correspond to the lengths of the strings of $S$ sorted in suffix lexicographic order ($''CA'',''B'',''BC'',''ABC''$).
\end{itemize}
 %on ferme l'environnement figure

\section{Construction algorithm}\label{sec:construction_details}
We now describe more in detail how each component of our representation is built: 
\paragraph{Construction of the transition dictionary}
The transition dictionary can be constructed by a simple DFS traversal of the suffix tree. Essentially each leaf of the suffix tree represents one or more suffixes of strings in $R$ (there could be multiple identical suffixes) which correspond to prefixes of strings in $S$ (recall that $R$ contains the strings of $S$ written in reverse order). The transition dictionary is then built in the following way: for each leaf of rank $i$, we have a list of suffixes of strings in $R$. Each suffix is represented by a pair $(string\mathunderscore pointer,\mbox{suf\mathunderscore pos})$. Thus for each suffix to which corresponds a pair $(string\mathunderscore pointer,\mbox{suf\mathunderscore pos})$, we add to the dictionary the pair $(c,i)$ where $c=string\mathunderscore pointer[\mbox{suf\mathunderscore pos}-1]$ at the condition that $\mbox{suf\mathunderscore pos}\geq 1$ (the suffix does not represent an element of $R$). In other words for each prefix $p$ of an element $x\in S$ such that $p\neq x$, $c$ is the successor of $p$ in $x$. 
 \\
Then we can sort the pairs using radix sort (and remove duplicates) which takes $O(n)$ time (recall that we have assumed that $\sigma\leq n$) and finally build the succinct indexable dictionary of \cite{RRR02} on the sorted output. 
\paragraph{Building of the failure tree}
For building the failure tree, we do a DFS traversal of the suffix tree and transform it progressively. When traversing a node $t$ having a parent $r$ (in the special case where $t$ is root we simply recursively traverse all of its children in lexicographic order), we check if the first child of $t$ (we note that child by $f$) is labeled with character $\#$ (which means that the child $f$ is a leaf representing a string of $P$ which is a suffix of all strings representing all other descendants of node $n$). If this is the case, we eliminate the child $f$. If the child $f$ is not labeled with $\#$, we eliminate the node $t$ and attach all its children in their original order in the list of children of node $r$ (recall that $r$ is the parent of node $t$) at the position where the node $t$ was attached (between the left sibling and right sibling of node $t$). Then we recursively process all the children of $t$ in lexicographic order (including the child $f$). 

\paragraph{Building of the report tree}
The report tree can be built from the failure tree by flattening progressively the tree in a top-down traversal. More precisely, we traverse the failure tree in DFS order. Each time we encounter a node $n$ corresponding to a string $p$, we first check whether $p\in S$ (this can easily be done by checking whether the list of pairs associated with $n$ contains a single pair $(string\mathunderscore pointer,suf\mathunderscore pos)$ with $position=0$). If it is the case, we keep it and recursively traverse all of its children. Otherwise ($p\notin S$), we attach all the children of $n$ as children of the parent of $n$ in the same order at the position where the node $n$ was attached (between right and left siblings of $n$) and continue to scan the children of $n$ recursively doing the same thing.

\paragraph{Building of the state dictionary}
The state dictionary can also be built in linear time. For that we can simply do a DFS traversal of the report tree during which a counter initialized at zero is incremented each time we encounter a node. Each time we encounter an internal node, we put the counter value in a temporary dictionary (the value of the counter before traversing a given node is exactly the value of the state corresponding to that node). This first step takes linear time. Building the compressed representation of $\cite{RRR02}$ on the array also takes randomized expected linear time. 

\paragraph{Building of the string length store}
The string length store can trivially be built in $O(n)$ time. For that we traverse the strings of $S$ in suffix lexicographic order and store the length of each string in a temporary array. Then, we can encode the temporary array using Elias-Fano encoding. 

\section{Compressed representation}
\label{compressed_aho}
We now give a proof of theorem~\ref{compressed_theorem}. The space usage of the transition dictionary can be reduced from $m(\log\sigma+1.443+o(1))$ to $m(H_0+1.443+o(1))$, where $H_0$ is the entropy of the characters appearing in the trie representation of the set $S$ (or equivalently the characters appearing in the $next$ transitions of the automaton). For that we will use $\sigma$ indexable dictionaries instead of a single one. Each dictionary corresponds to one of the characters of the alphabet. That is a pair $(c,state)$ will be stored in the dictionary corresponding to character $c$ (we note that dictionary by $I[c]$).  Additionally we store a table $T[0..\sigma-1]$. For each character $c$ we set $T[c]$ to the rank of character $c$ (in suffix-lexicographic order) relatively to the set $P$ (that is the number of strings in the set $P$ which are smaller than the string $''c''$ in the suffix lexicographic order). Let $Y$ be the set of pairs to be stored in the transition dictionary. The indexable dictionary $I[c]$ will store all values $state_i$ such that $(c,state_i)\in Y$. Thus the number of elements stored in $I[c]$ is equal to the number of $next$ transitions labeled with character $c$.
\\
Now the target state for a transition pair $(c,state)$ is obtained by $T[c]+rank_{I[c]}(state)$, where $rank_{I[c]}(state)$ is the rank operation applied on the dictionary $I[c]$ for the value $state$. Let's now analyze the total space used by the table $T$ and by the indexable dictionaries. The space usage of table $T$ is $\sigma\log m\leq m^{\epsilon}\log m=o(m)$. An indexable dictionary $I[c]$ will use at most $t_c(\log(m/t_c)+1.443+o(1))$ bits , where $t_c$ is the number of transitions labeled with character $c$. Thus the total space used by all indexable dictionaries is $\sum_{0\leq c<\sigma}t_c(\log(m/t_c)+1.443+o(1))=m(H_0+1.443+o(1))$. Thus the total space used by the table $T$ and the indexable dictionaries is $m(H_0+1.443+o(1))$.
\end{document}